\documentclass[10pt,onecolumn,aps,prd,preprintnumbers,showpacs,superscriptaddress,nofootinbib,amsmath,amssymb,floats,floatfix,showkeys,notitlepage,longbibliography]{revtex4-1}

\usepackage{orcidlink}
\usepackage{comment}
\usepackage{lipsum}
\usepackage{graphicx}
\usepackage{subfigure}
\usepackage{palatino}
\usepackage{sans}
\usepackage{hyperref}
\hypersetup{colorlinks=true,linkcolor=blue,urlcolor=blue,citecolor=blue}
\usepackage[toc,page]{appendix}
\usepackage[normalem]{ulem}
\usepackage{adjustbox}
\usepackage{latexsym}
\usepackage{amsmath}
\usepackage{amssymb}
\usepackage{amsfonts}
\usepackage{dcolumn}
\usepackage{bm}
\usepackage{tikz}
\usetikzlibrary{decorations.pathmorphing}
\usepackage{bigints}
\usepackage{array,tabularx,multirow,booktabs}
\usepackage[tracking=true]{microtype}
\usepackage{soul} 
\SetTracking{}{500}
\SetTracking{encoding={*}, shape=sc}{40}
\UseRawInputEncoding 
\allowdisplaybreaks

\usepackage[utf8]{inputenc}
\usepackage{algorithm}
\usepackage{algorithmicx}
\usepackage{algpseudocode}
\usepackage{xcolor} 

\begin{document} \sloppy

\title{Acceleration Radiation from Derivative-Coupled Atoms Falling in Modified Gravity Black Holes}

\author{Reggie C. Pantig \orcidlink{0000-0002-3101-8591}} 
\email{rcpantig@mapua.edu.ph}
\affiliation{Physics Department, School of Foundational Studies and Education, Map\'ua University, 658 Muralla St., Intramuros, Manila 1002, Philippines.}

\author{Ali \"Ovg\"un \orcidlink{0000-0002-9889-342X}}
\email{ali.ovgun@emu.edu.tr}
\affiliation{Physics Department, Faculty of Arts and Sciences, Eastern Mediterranean University, Famagusta, 99628 North
Cyprus via Mersin 10, Turkiye.}

\begin{abstract}
The interaction of quantum detector models with fields in curved spacetimes provides fundamental insights into phenomena such as Hawking and Unruh radiation. While standard models typically assume a minimal coupling between the detector and the field, physically motivated derivative couplings, which are sensitive to field gradients, have been less explored, particularly in the context of modified gravity theories. In this paper, we develop a general framework to analyze the acceleration radiation from a two-level atomic detector with a derivative coupling undergoing a radial geodesic infall into a generic static, spherically symmetric black hole. We derive a general integral expression for the excitation probability and apply it to two distinct spacetimes. For an extended uncertainty principle (EUP) black hole, we demonstrate that the detector radiates with a perfect thermal spectrum at the precise Hawking temperature, reinforcing the universality of this phenomenon. For a black hole solution in a Ricci-coupled Bumblebee gravity model, the radiation is also thermal. Still, its temperature is modified in direct correspondence with the theory's Lorentz-violating parameters, consistent with the modified Hawking temperature. Furthermore, we demonstrate that derivative coupling results in a significantly enhanced entropy flux compared to minimal coupling models. Our results establish acceleration radiation as a sensitive probe of near-horizon physics and demonstrate that this phenomenon can provide distinct observational signatures to test General Relativity (GR) and alternative theories of gravity in the strong-field regime.
\end{abstract}


\maketitle

\section{Introduction} \label{intro}
The intersection of General Relativity (GR) and Quantum Mechanics (QM) remains one of the most profound frontiers in theoretical physics. In this grand endeavor, black holes serve as unparalleled cosmic laboratories, where the possible link between spacetime curvature and quantum fields manifests in its most extreme forms. It was in this arena where Jacob Bekenstein proposed (in 1972) that black holes possess entropy proportional to their horizon area, drawing analogies from the second law of thermodynamics and information theory \cite{Bekenstein:1972tm}, which sets the stage for quantum corrections in black holes leading to the foundations of black hole thermodynamics \cite{Bekenstein:1973ur}. Then, Hawking famously predicted that black holes are not truly black, but radiate thermally as a consequence of quantum vacuum fluctuations near their event horizon \cite{Hawking:1974rv,Hawking:1975vcx}. A related phenomenon, the Unruh effect, posits that an accelerating observer will perceive the vacuum as a thermal bath, revealing an intimate connection between acceleration, thermodynamics, and the quantum nature of spacetime \cite{Unruh:1976db,Unruh:1982ic,Unruh:1983ms,Crispino:2007eb}. While not explicitly named, the prototype of the Unruh-DeWitt detector emerged as a two-level system linearly coupled to a massless scalar field, used to compute transition rates in Rindler coordinates. This work established the detector as a tool to measure particle content near black hole horizons, where surface gravity mimics acceleration. Eventually, Bryce DeWitt formalized the detector model, now known as the Unruh-DeWitt detector, in his 1977 work on quantum field theory in curved spacetimes \citep{DeWitt:1975ys}. He described a point-like quantum system moving along a worldline, coupled to a massless scalar field via a monopole interaction. DeWitt computed the detector’s response function in Schwarzschild spacetime, showing that it registers thermal radiation for static observers near a black hole, consistent with Hawking radiation. This formalized the detector’s role in probing the Boulware \cite{Boulware:1974dm} and Hartle-Hawking vacua \cite{Hartle:1976tp}. The Unruh-DeWitt detector has been instrumental in advancing our understanding of quantum field theory in black hole spacetimes \cite{Takagi:1986kn,Srednicki:1993im,Fuentes-Schuller:2004iaz,Crispino:2007eb,Costa:2008st,Acquaviva:2011vq,Wang:2014uoa,Tjoa:2018xre,Camblong:2020pme,Gallock-Yoshimura:2021yok,Scully:2022pov,Chen:2023uxl}.

The vast majority of studies have employed a standard "minimal coupling" interaction, where the detector's monopole moment couples directly to the field amplitude \cite{Martin-Martinez:2012ysv}. While this approach has been remarkably successful, it represents only the simplest possible interaction. A more physically nuanced picture can be explored through "derivative couplings," where the detector interacts with the spacetime gradient of the field \cite{Martin-Martinez:2014qda}. Such couplings are not merely a mathematical curiosity; they arise naturally in effective field theories and are directly analogous to how physical forces, such as electromagnetism, are mediated by field strengths (i.e., derivatives of the potential). A detector with derivative coupling is sensitive not just to the field's presence, but to its local dynamics, making it a more discerning probe of the quantum vacuum's structure \cite{Lee:2012tvc}. In 2014, a study examined a derivative-coupled detector in 1+1 dimensions, deriving regulator-free expressions for transition probabilities and demonstrating sudden yet finite responses during the onset of the Hawking-Unruh effect, though primarily in Hartle-Hawking and Unruh vacua; extensions to Boulware-like states highlighted the vacuum's regularity issues at horizons without the infrared problems of non-derivative models \cite{Juarez-Aubry:2014jba}. Concurrently, Louko investigated a two-level detector coupled to either the field or its proper time derivative in a (1+1)-dimensional Minkowski setup with a Rindler firewall, analogous to the Boulware singularity, finding that the response remains finite across the horizon, with deviations from the vacuum scaling as $\ln(|\omega|)$ for large energy gaps $\omega$, thus suggesting that firewalls do not necessarily induce pathological detector behavior \cite{Louko:2014aba}. Subsequent studies have built on these foundations to distinguish black holes from horizonless exotic compact objects (ECOs), where the Boulware vacuum remains well-defined without horizon singularities \cite{Holdom:2020uhf,Tjoa:2022oxv}. The study of such couplings has revealed richer phenomenology, including non-trivial dependencies on the detector's trajectory and the potential for enhanced energy exchange \cite{Hsiang:2021lxp}. Throughout this work, we assume the scalar field to be in the Boulware vacuum state, which corresponds to the vacuum perceived by a static observer at infinity and is the natural choice for an isolated, non-radiating black hole.

More recently, a powerful quantum optics approach, pioneered by Scully and collaborators, has provided new insights by framing the problem in terms of an infalling atom interacting with quantized field modes, leading to the concept of "acceleration radiation" \cite{Ben-Benjamin:2019opz,Scully:2017utk,Svidzinsky:2018jkp,Azizi:2020gff,Azizi:2021qcu,Azizi:2021yto}. This framework has proven particularly effective for analyzing the entropy associated with the emitted radiation. The near-horizon region of a black hole is precisely where quantum gravity effects might become manifest, potentially altering the spacetime structure from the predictions of classical GR \cite{Blommaert:2020yeo}. Therefore, analyzing how quantum radiation phenomena are affected in these alternative theories provides a unique avenue to test fundamental physics \cite{coutant:tel-00747874}. A variety of recent investigations have highlighted novel aspects of acceleration radiation and its entropy in modified gravitational and quantum frameworks. Lorentz-violating effects have been shown to imprint observable signatures on the acceleration radiation spectrum \cite{Tang:2025eew}, while equivalence principle considerations and quantum corrections introduce additional entropy contributions for atoms falling into nonclassical black hole backgrounds \cite{Sen:2022tru,Sen:2022cdx}. Extensions to braneworld models reveal enhanced horizon-brightened radiation due to extra-dimensional influences \cite{Das:2023rwg}. Charged black hole analyses uncover charge-dependent and inverse logarithmic entropy corrections under renormalization group improvements \cite{Jana:2024fhx,Jana:2025hfl}.  In a recent study, the entropy of horizon-brightened acceleration radiation (HBAR) for atoms infalling into a generalized uncertainty principle (GUP)–corrected Schwarzschild black hole was computed, revealing subtle equivalence principle violations at the quantum level \cite{Ovgun:2025isv}. Investigations of derivative coupling within a quantum optics framework have further demonstrated how nonminimal detector–field interactions alter both the spectrum and entropy of horizon-brightened acceleration radiation from Schwarzschild's black hole \cite{Das:2025rzz}. Moreover, cosmological factors such as dark energy modify nonthermal atomic radiation profiles near black holes \cite{Bukhari:2022wyx}, and acceleration radiation emerges as a potential dark matter probe \cite{Bukhari:2023yuy}. Finally, Lorentz violation has been further recognized to affect radiative transitions and associated entropy in spherically symmetric black hole settings \cite{Rahaman:2025grm, Rahaman:2025mrr}.

The quantum optics framework also clarifies an important physical question: why should a "force-free" detector following a geodesic path radiate at all? The answer lies in the distinction between kinematic acceleration and the detector's acceleration relative to the quantum field modes. While a geodesic observer has zero kinematic acceleration, the Boulware vacuum is a state defined by modes that are static with respect to observers at infinity. As the detector falls freely, it experiences a non-zero proper acceleration relative to these field modes, which are effectively held up by the black hole's gravitational field \cite{Scully:2017utk}. It is this relative acceleration that induces the emission, justifying the term "acceleration radiation" even for a detector on a geodesic trajectory.

While recent work has shown that derivative coupling between a detector and field can significantly modify the static-detector radiation spectrum, the behavior of such non-minimal interactions along realistic, geodesic free-fall trajectories has, until very recently, remained unexplored. Concurrently with our present investigation, a complementary study by Das et al. \cite{Das:2025rzz} analyzed this phenomenon for the specific case of the Schwarzschild black hole. In this paper, we address the topic from a more fundamental standpoint by developing a general framework for computing acceleration radiation from a detector with derivative coupling undergoing radial, geodesic free-fall into any static, spherically symmetric black hole. Our primary contribution is a general integral expression for the excitation probability that can serve as a versatile tool for probing various black hole geometries. To demonstrate its utility, we first apply our formalism to an EUP black hole spacetime \cite{Mureika:2018gxl} to illustrate the universality of the resulting thermal spectrum. We then extend the analysis to a black hole solution in Ricci-coupled Kalb-Ramond Bumblebee gravity to explore how fundamental symmetry breaking alters both the spectrum and entropy of horizon-brightened acceleration radiation \cite{Belchior:2025xam}.

The paper is structured as follows: In Section \ref{sec2}, we construct the interaction Hamiltonian for the derivative coupling model. In Section \ref{sec3}, we derive the general expression for the detector's excitation probability. In Sections \ref{sec4} and \ref{sec5}, we apply this formalism to the EUP black hole and Bumblebee black hole spacetimes, respectively, and analyze the resulting radiation spectra. In Section \ref{sec6}, we discuss the implications of our results for the entropy of acceleration radiation. We conclude in Section \ref{conc} with a summary of our findings and a discussion of future research directions. Throughout this paper, we used geometrized units by setting $G=c=k_B=\hbar=1$. We also used the metric signature $(+,-,-,-)$.

\section{Interaction Hamiltonian with Derivative Coupling} \label{sec2}
Our analysis is based on a quantum optics approach, where we use first-order, time-dependent perturbation theory to calculate the excitation probability of the detector. We model the detector as a point-like two-level quantum system (or qubit), which serves as a simplified model for a real atom. It possesses a ground state $\lvert \text{g} \rangle$ and an excited state $\lvert \text{e} \rangle$, separated by an energy gap $\omega$, which represents the energy required for a transition from the ground to the excited state. The detector's internal dynamics are governed by the free Hamiltonian $H_D = \omega \, \hat{\sigma} \hat{\sigma}^\dagger$, where $\hat{\sigma}$ and $\hat{\sigma}^\dagger$ are the standard lowering and raising operators for the two-level system, respectively. The use of the term "atom" throughout this paper is a convention adopted from the quantum optics literature to provide a clear physical analogy \cite{Scully:2017utk}.

The detector is coupled to a massless, real scalar field, $\phi$, which we assume to be in the Boulware vacuum state. This state is defined by selecting only positive frequency modes with respect to the asymptotic time coordinate $t$. It corresponds to the vacuum state perceived by a static observer at infinity and is considered the most natural state for describing an isolated, non-radiating black hole. The interaction between the detector and the field is described by an interaction Hamiltonian, $\hat{V}$. In the interaction picture, this Hamiltonian has the form
\begin{equation}
    \hat{V}(\tau) = \hbar g \left[\hat{a}_\nu e^{-i\nu [t(\tau) - r_*(\tau)]} + \text{H.c.} \right] \hat{\mu}(\tau),
\end{equation}
where $g$ is a small coupling constant, signifying that we will be working within first-order perturbation theory, and $\hat{\mu}(\tau)$ is the detector's monopole operator, given by $\hat{\mu}(\tau) = \hat{\sigma} e^{-i\omega \tau} + \hat{\sigma}^\dagger e^{i\omega \tau}$ \cite{Bukhari:2023yuy}.

Instead of coupling to the field $\phi$, we couple to its proper-time derivative along the atom’s worldline. This leads to the modified interaction Hamiltonian:
\begin{equation} \label{e_hamil_der}
	\hat{V}_\text{der}(\tau) = \hbar \tilde{g} \left[ \frac{d}{d\tau} \hat{\phi}(x(\tau)) \right] \left( \hat{\sigma} e^{-i\omega \tau} + \text{H.c.} \right),
\end{equation}
where we define $\tilde{g} = g/\omega$ for dimensional consistency. The interaction term is manifestly covariant and represents a physical observable. The proper time derivative can be written as a Lorentz scalar, $d\phi/d\tau = u^\mu \partial_\mu \phi$, where $u^\mu = dx^\mu/d\tau$ is the detector's four-velocity and $\partial_\mu\phi$ is the gradient of the scalar field. This ensures that the physical interaction is independent of the chosen coordinate system. In natural units where action is dimensionless, the Hamiltonian must have dimensions of energy. The scalar field $\phi$ has dimensions of energy in four-dimensional spacetime. The proper-time derivative, $d/d\tau$, introduces an additional dimension of energy. Consequently, for the overall interaction term to have the correct units, the coupling constant $\tilde{g}$ must have dimensions of inverse energy. Tying it to the detector's own characteristic energy scale, $\omega$, is a physically intuitive way to introduce the required dimensional factor. This ensures that the interaction strength is naturally related to the internal properties of the detector itself.

To make it explicit in terms of field modes, write:
\begin{equation}
	\hat{\phi}(x(\tau)) = \sum_\nu \left[ \hat{a}_\nu \psi_\nu(x(\tau)) + \hat{a}_\nu^\dagger \psi_\nu^*(x(\tau)) \right].
\end{equation}
Here, the sum is over the field modes, which are labeled by their frequency $\nu > 0$. Then, the total interaction becomes:
\begin{equation} \label{e_tot_interaction}
	\hat{V}_\text{der}(\tau) = \hbar \tilde{g} \sum_\nu \left[ \hat{a}_\nu \frac{d\psi_\nu(x(\tau))}{d\tau} + \hat{a}_\nu^\dagger \frac{d\psi_\nu^*(x(\tau))}{d\tau} \right] \left( \hat{\sigma} e^{-i\omega \tau} + \text{H.c.} \right).
\end{equation}
The physical basis for such a coupling is robust. In the framework of effective field theory, derivative couplings naturally arise as higher-order operators. They can represent interactions that become dominant at specific energy scales or that are mediated by heavier fields not explicitly included in the low-energy model. For example, in quantum electrodynamics, the physical electric and magnetic forces are determined by the field-strength tensor $F_{\mu\nu} = \partial_\mu A_\nu - \partial_\nu A_\mu$, which involves derivatives of the vector potential $A^\mu$. An interaction with $F_{\mu\nu}$ is thus sensitive to the gradients of the potential. In a similar spirit, a derivative coupling of the form $u^\mu \partial_\mu \phi$ makes the detector sensitive to the spacetime gradient of the scalar field, $\partial_\mu \phi$, as measured in its own rest frame (since $u^\mu \partial_\mu = d/d\tau$). This contrasts with the standard minimal coupling, which only probes the field's local amplitude, $\phi$.

Given that the field mode $\psi_\nu$ as \cite{Scully:2017utk}
\begin{equation}
	\psi_\nu(t, r_*) = e^{i \nu (t - r_*)},
\end{equation}
and that along the atomic trajectory $x^\mu(\tau)$, the mode function becomes
\begin{equation} \label{e_field_mode}
	\psi_\nu[x(\tau)] = e^{i\nu[t(\tau) - r_*(\tau)]}.
\end{equation}
We note that these plane-wave modes are not canonically normalized. Following the convention of the quantum optics approach \cite{Scully:2017utk}, any normalization constants are absorbed into the definition of the coupling constant $g$. This does not affect the spectral shape of the final excitation probability. Then,
\begin{equation}
	\frac{d}{d\tau} \psi_\nu[x(\tau)] = i\nu \left( \frac{dt}{d\tau} - \frac{dr_*}{d\tau} \right) e^{i\nu[t(\tau) - r_*(\tau)]}.
\end{equation}
Therefore, the new interaction Hamiltonian becomes
\begin{equation} \label{e_hamil_der2}
\hat{V}_\text{der}(\tau) = i\hbar \tilde{g} \sum_\nu \nu \left( \frac{dt}{d\tau} - \frac{dr_*}{d\tau} \right) \left[ \hat{a}_\nu e^{i\nu(t - r_*)} - \hat{a}_\nu^\dagger e^{-i\nu(t - r_*)} \right] \left( \hat{\sigma} e^{-i\omega \tau} + \hat{\sigma}^\dagger e^{i\omega \tau} \right).
\end{equation}
In a physical sense, this coupling accounts for the rate of change of the field as experienced along the atom’s worldline. It aligns with Unruh-DeWitt detector models with derivative coupling.

A direct comparison with the standard Unruh-DeWitt Hamiltonian, which involves a minimal coupling to the field amplitude $\phi$ itself, is instructive. Minimal coupling effectively tests the local amplitude of quantum vacuum fluctuations. In contrast, the derivative coupling model probes the dynamics of these fluctuations as perceived by the detector. The detector's excitation becomes sensitive not just to the presence of a field mode, but to the rate at which the phase of that mode changes along its worldline. This sensitivity to field gradients is particularly relevant in the highly non-inertial environment of a black hole infall, where the extreme gravitational field leads to significant Doppler shifts and compression of the field modes. It is important to emphasize that this interaction is fundamentally local, as the derivative is evaluated at the detector's instantaneous position $x(\tau)$. In our perturbative approach, we directly integrate this local interaction vertex over the detector's worldline. Consequently, the potential non-local effects that can arise from boundary terms via integration by parts do not appear in this framework.

\section{The probability amplitude and excitation probability} \label{sec3}
We now apply our formalism to the specific physical scenario of an infalling detector. We assume the detector starts at rest at spatial infinity and follows a purely radial, timelike geodesic into a static and spherically symmetric black hole. The spacetime is described by the line element
\begin{equation} \label{e_metric}
    ds^2 = f(r) dt^2 - \frac{1}{f(r)} dr^2 - r^2 (d\theta^2 + \sin^2\theta d\phi^2), 
\end{equation}
where the metric function $f(r)$ encapsulates the gravitational influence of the central object, whose asymptotic (ADM) mass is $M$. For a particle starting at rest at infinity, its conserved energy per unit mass, $E$, is equal to 1. This classical parameter $E$, which defines the geodesic trajectory, is entirely independent of the detector's internal quantum energy gap $\omega$. The geodesic equations of motion for such a radial trajectory ($\dot{\theta} = \dot{\phi} = 0$), derived from the normalization condition $g_{\mu\nu}u^\mu u^\nu = 1$, are then
\begin{equation}
    \frac{dr}{d\tau} = -\sqrt{E^2 - f(r)} = -\sqrt{1 - f(r)}.
\end{equation}
The coordinate time also evolves as
\begin{equation}
    \frac{dt}{d\tau} = \frac{E}{f(r)} = \frac{1}{f(r)}.
\end{equation}
Also recall that in terms of tortoise coordinate,
\begin{equation}
    \frac{dr_*}{dr} = \frac{1}{f(r)} \Rightarrow \frac{dr_*}{d\tau} = \frac{dr_*}{dr} \frac{dr}{d\tau} = \frac{-\sqrt{1 - f(r)}}{f(r)}.
\end{equation}

The probability amplitude is given by first-order perturbation theory,
\begin{equation}
    \mathcal{A} = (i\hbar)^{-1}\int \langle f | \hat{V}_\text{der}(\tau) | i \rangle d\tau. 
\end{equation}
The initial state is $|i\rangle = |0,g\rangle$ (vacuum field, ground-state atom), and the final state is $|f\rangle = |1_{\nu},e\rangle$ (one particle in the field, excited-state atom). From the full interaction Hamiltonian in Eq. \eqref{e_tot_interaction}, the only term that contributes to this excitation process is the one containing the creation operators $\hat{a}_{\nu}^\dagger$ and $\hat{\sigma}^\dagger$. The corresponding matrix element is:
\begin{align}
    \langle 1_{\nu},e | \hat{V}_\text{der}(\tau) | 0,g \rangle &= \langle 1_{\nu},e | \hbar \tilde{g} \left( \hat{a}_\nu^\dagger \frac{d\psi_\nu(x(\tau))}{d\tau} \right) \left( \hat{\sigma}^\dagger e^{i\omega \tau} \right) | 0,g \rangle  \nonumber \\
    &= \hbar \tilde{g} \frac{d\psi_\nu(x(\tau))}{d\tau} e^{i\omega \tau} \langle 1_{\nu},e | \hat{a}_\nu^\dagger \hat{\sigma}^\dagger | 0,g \rangle \nonumber \\
    &= \hbar \tilde{g} \frac{d\psi_\nu(x(\tau))}{d\tau} e^{i\omega \tau}.
\end{align}
Substituting the explicit form of the mode derivative from Eq. \eqref{e_hamil_der2}, this becomes:
\begin{equation}
    \langle 1_{\nu},e | \hat{V}_\text{der}(\tau) | 0,g \rangle = -i \hbar \tilde{g} \nu \left( \frac{dt}{d\tau} - \frac{dr}{d\tau} \right) e^{-i\nu[t(\tau) - r_*(\tau)]} e^{i\omega \tau}. \nonumber
\end{equation}
The probability amplitude for excitation is the integral of this matrix element. Since we are ultimately interested in the probability $P=|\mathcal{A}|^2$, the overall phase of the amplitude and the sign in the complex exponential do not affect the final result. We can therefore write the core integral for the amplitude as:
\begin{equation} \label{e_A_der}
    \mathcal{A}_{\text{exc}}^{\text{(der)}} \propto \tilde{g} \nu \int d\tau \left( \frac{dt}{d\tau} - \frac{dr}{d\tau} \right) e^{i\nu[t(\tau) - r_*(\tau)]} e^{i\omega \tau(r)}.
\end{equation}

Therefore, along with the simple chain-rule $d\tau = \frac{d\tau}{dr} dr$, and after combining terms, Eq. \eqref{e_A_der} becomes
\begin{equation}
\mathcal{A}_{\text{exc}}^{\text{(der)}} = i \hbar \tilde{g} \nu \int^{r=\infty}_{r=r_{\rm h}^+} dr \, F(r) \, e^{i\nu[t(r) - r_*(r)]} e^{i\omega \tau(r)},
\end{equation}
where $F(r)$ is the geometric prefactor
\begin{equation} \label{e_prefactor}
      F(r) = \frac{1 + \sqrt{1-f(r)}}{f(r)\sqrt{1-f(r)}}.
\end{equation}
Note that the phase of the integrand, $\nu[t(r) - r_*(r)] + \omega \tau(r)$, is determined solely by the detector's worldline and is identical to the phase that appears in the standard minimal coupling model. The novelty of the derivative coupling is captured entirely by the new geometric prefactor, $F(r)$. Since the thermal properties of the spectrum are governed by the behavior of this phase near the horizon, we anticipate a thermal spectrum, but with an overall amplitude that is modified by this new prefactor. The excitation probability is then defined as
\begin{eqnarray} \label{e_Pex}
P_{\text{exc}}^{\text{(der)}} = \hbar^{-2} \left| \mathcal{A}_{\text{exc}}^{\text{(der)}} \right|^2= \tilde{g}^2 \nu^2 \left| \int^{r=\infty}_{r=r_{\rm h}^+} dr \, F(r) \, e^{i\nu[t(r) - r_*(r)]} e^{i\omega \tau(r)} \right|^2,
\end{eqnarray}
As we see, we obtained a new and more complex geometric prefactor inside the integral due to the derivative coupling. In the following sections, we apply it to the EUP line element and in the black hole influenced by the Bumblebee parameter.

The emergence of the new, more complex geometric prefactor, $F(r)$, is the central consequence of using a derivative coupling model. It's crucial to understand its physical significance. This term is not merely a mathematical artifact but a direct reflection of how the detector's interaction is shaped by the spacetime curvature: First, the appearance of $f(r)$ in the denominator is directly tied to gravitational time dilation ($dt/d\tau = 1/f(r)$) and the definition of the tortoise coordinate $dr_*$. Second, the term $\sqrt{1-f(r)}$ originates from the detector's proper radial velocity ($dr/d\tau$) and is therefore directly related to its proper acceleration. The combination of these factors shows that the derivative coupling makes the detector's response sensitive not just to its location, but dynamically sensitive to its state of motion through the curved background. The interaction strength is amplified precisely where gravitational effects and acceleration are most extreme, namely, near the event horizon.

From Eq. \eqref{e_Pex}, the expressions for $t(r)$, $r_*(r)$, and $\tau(r)$ are of crucial importance. With the help of the geodesic equations, we find the expressions for these quantities as
\begin{align} \label{e_geo}
    t(r) &= -\int \frac{dr}{\sqrt{1-f(r)}}, \nonumber \\
    r_*(r) &= \int \frac{dr}{f(r)}, \nonumber \\
    \tau(r) &= -\int \frac{dr}{f(r)\sqrt{1-f(r)}}.
\end{align}
\begin{figure}
{\includegraphics[width=0.6\textwidth]{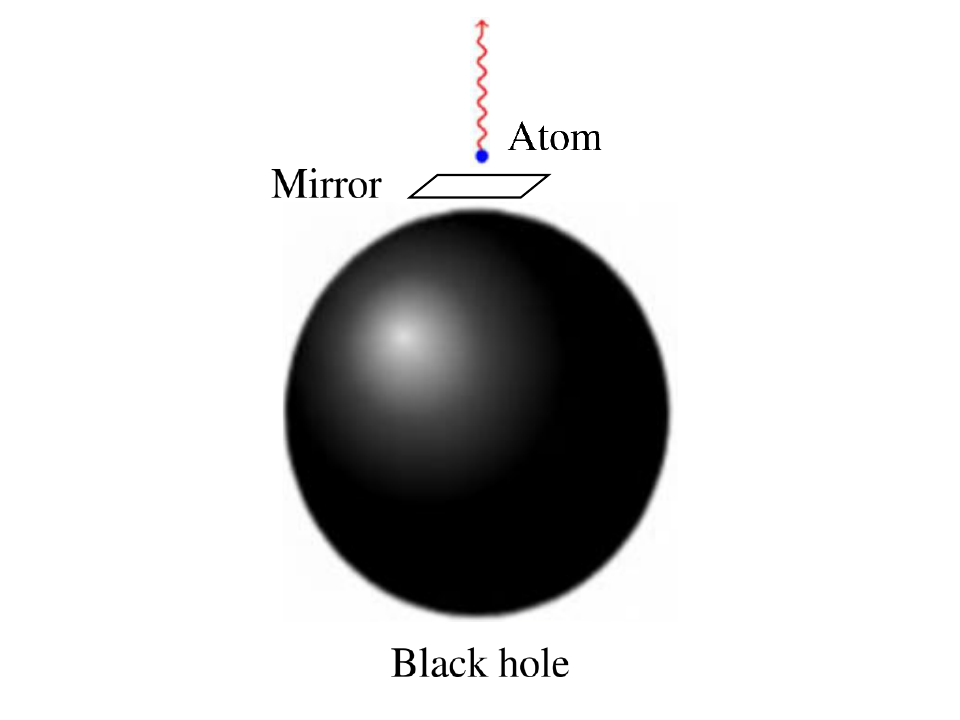}} 
    \caption{Conceptual illustration of acceleration radiation from an atom near a black hole. A reflective boundary, or "mirror," is held at a fixed position outside the event horizon. This boundary establishes a defined Boulware-like vacuum state for the infalling atom, effectively shielding it from pre-existing Hawking radiation. As the atom plunges through this vacuum, the immense proper acceleration it experiences relative to the quantum field modes induces the emission of a photon, a phenomenon known as acceleration radiation.}
    \label{fig1}
\end{figure}

\section{Example 1: Extended Uncertainty Principle Black Hole} \label{sec4}
In this analysis, we consider a black hole model proposed by Mureika \cite{Mureika:2018gxl} that incorporates the Extended Uncertainty Principle (EUP). The EUP is a phenomenological modification of the standard quantum uncertainty relation that introduces a correction term dependent on a new, large fundamental distance scale, $L_{*}$ \cite{Bambi:2007ty}. The principle is expressed as:
\begin{equation} \label{e1}
    \Delta x \Delta p \geq \frac{\hbar}{2} \left( 1+ \frac{\alpha\Delta x^2}{L_{*}^2} \right) \nonumber
\end{equation}
(restoring $\hbar$ for clarity), where $\alpha$ is the EUP parameter, generally assumed to be unity. We acknowledge that this relation is a phenomenological model and its non-manifestly covariant form represents a departure from the standard canonical framework. Our aim in this paper is not to provide a first-principles derivation of this principle, but rather to employ the resulting black hole metric as a consistent, quantum-corrected spacetime to test our acceleration radiation formalism.

To apply this to a black hole, the model draws upon the quantum-N portrait, which posits that a black hole can be viewed as a composite system of $N$ soft gravitons \cite{Dvali:2011aa}. While this concept is a subject of ongoing research, in this context, the event horizon is conceptualized as a boundary that confines a collection of $N$ gravitons. The uncertainty in the position of these gravitons is taken to be the size of the event horizon, $\Delta x \sim 2m$, where $M$ is the mass of the black hole. This leads to a modified momentum uncertainty for each graviton:
\begin{equation} \label{e2}
    \Delta p_{g} \sim \frac{\hbar }{2M} \left ( 1 + \frac{4\alpha M^{2}}{L_{*}^{2}} \right ).
\end{equation}
Assuming the mass of a single graviton is $\frac{\hbar}{2M}$, the total mass of the black hole can be approximated by $M \sim N \frac{\hbar}{2M}$. Consequently, the total momentum uncertainty, $\Delta P$, for all the confined gravitons is:
\begin{equation} \label{e3}
    \Delta P \sim M \left ( 1 + \frac{4\alpha M^{2}}{L_{*}^{2}} \right ).
\end{equation}
The stress-energy tensor includes EUP corrections, the ADM mass is modified. This EUP-corrected mass, $M_{\text{EUP}}$, is equivalent to the total momentum uncertainty, $\Delta P$:
\begin{equation} \label{e5}
    M_{\text{EUP}} = M \left ( 1 + \frac{4\alpha M^{2}}{L_{*}^{2}} \right ).
\end{equation}
This new mass term allows for the formulation of an EUP-inspired Schwarzschild metric function:
\begin{equation} \label{e6}
    f(r) = 1 - \frac{2M}{r} \left ( 1 + \frac{4\alpha M^{2}}{L_{*}^{2}}\right ).
\end{equation}
The location of the event horizon, $r_{\rm h}^{\rm EUP}$, is determined by setting $f(r)=0$, which yields:
\begin{equation} \label{e7}
    r_{\rm h}^{\rm EUP} = 2M + \frac{8\alpha M^{3}}{L_{*}^{2}}.
\end{equation}
The new scale $L_*$ introduces a model-dependent relationship between the black hole's mass and the quantum gravity scale. As investigated in \cite{Mureika:2018gxl, Dabrowski:2019wjk}, for values of $\alpha=1$ and $L_*$ in the range of $10^{12}$ m to $10^{14}$ m (potentially related to the Hubble scale), EUP effects become significant for supermassive black holes with masses from $10^9$ to $10^{11}$ solar masses. Additionally, there are potential connections between $L_*$ and the Hubble length scale, which could have broader implications for our understanding of galactic dynamics and cosmology \cite{Dabrowski:2019wjk}. For brevity, we want to rescale this horizon as $r_{\rm h}^{\rm EUP} = 1$. This is a standard technique to make the subsequent calculations and the final spectral form dimensionless. All dimensionful quantities, such as frequency $\nu$, are effectively being measured in units of the black hole's mass. As such, Eq. \eqref{e_Pex} becomes
\begin{align} \label{P_ex_inc}
    P_{\text{exc}}^{\text{(der)}} =g^2 \nu^2 \left|\int_{1}^\infty dr \frac{r}{\sqrt{r}-1} e^{i\nu\left[\left(t-r_* \right) \right]} e^{i \omega \tau} \right|^2,
\end{align}
where
\begin{align} \label{e_tau_t_r_*_res}
	\tau &= -\frac{2 r^{3/2}}{3} + C_1, \nonumber \\
	t &= -2\sqrt{2}-\frac{2 r^{3/2}}{3} + \ln\left( 1 + \sqrt{r} \right) - \ln\left( 1 - \sqrt{r} \right) + C_2, \nonumber \\
	r_* &= r + \ln\left( r - 1 \right) + C_3.
\end{align}
The integration constants $C_1$, $C_2$, and $C_3$ contribute only an overall constant phase to the probability amplitude and thus do not affect the final probability, which is the modulus squared of the amplitude. Upon substitution of these equations into Eq. \eqref{P_ex_inc}, and keeping in mind that the modulus of any complex number of the form $e^{i\theta}$ (where $\theta$ is a phase constant) is equal to $1$, then we find
\begin{equation}
	P_{\text{exc}}^{\text{(der)}} = g^2 \nu^2 \left|\int_{1}^\infty dr \frac{r}{\sqrt{r}-1} e^{-i \nu \Phi(r)} e^{-(2/3)i \omega r^{3/2}} \right|^2,
\end{equation}
where the phase function is
\begin{equation}
	\Phi(r) = \left[ 2\sqrt{r} + r + (2/3)r^{3/2} + 2\ln\left( \sqrt{r} - 1 \right) \right].
\end{equation}

Next, we apply some change in variables \cite{Scully:2017utk} to simplify the prefactor. The first one is $y = r^{3/2}$, where its differential is $dr =(2/3)y^{-1/3}dy$. Then,
\begin{equation}
    \frac{r \, dr}{\sqrt{r}-1} = \frac{2}{3}\frac{y^{1/3} dy}{\left(y^{1/3}-1\right)}.
\end{equation}
Then, the next change in variable is $x = (2/3)\omega(y-1)$. This gives
\begin{equation}
    \frac{y^{1/3}}{y^{1/3}-1} = \frac{\left(1 + \frac{3x}{2\omega}\right)^{1/3}}{\left(1 + \frac{3x}{2\omega}\right)^{1/3}-1}.
\end{equation}
The dominant contribution to the integral arises from the behavior of the integrand near the event horizon. We are therefore interested in its behavior near the coordinate singularity at the horizon, which corresponds to $r \rightarrow 1$ in our scaled coordinates. This limit implies $y \rightarrow 1$ and thus $x \rightarrow 0$. The expansion leads to
\begin{equation}
    \frac{\left(1 + \frac{3x}{2\omega}\right)^{1/3}}{\left(1 + \frac{3x}{2\omega}\right)^{1/3}-1} \approx \frac{2\omega}{x}.
\end{equation}
Then, after some strenuous algebra, we finally find
\begin{equation} \label{e_Pex2}
    P_{\text{exc}}^{\text{(der)}} = g^2\nu^2  \left| \left(2\omega\right)^{2i\nu} e^{-iA} \int_{0}^\infty dx \, x^{-1-2i\nu} e^{-iBx} \right|^2,
\end{equation}
where
\begin{equation}
	A = \frac{11\nu}{3} +\frac{2\omega}{3}, \quad B = 1+\frac{3\nu}{\omega}.
\end{equation}
Note that in this transition, the factor of $1/\omega$ from the change of measure $dy = (3/2\omega)dx$ is cancelled by a factor of $\omega$ that arises in the numerator of the approximation for the geometric prefactor.

The above is of the form
\begin{equation}
	\int_0^\infty dx \, x^{s - 1} e^{-i B x}, \quad \text{with} \quad s = -2i\nu, \, B = 1 + \frac{3\nu}{\omega}.
\end{equation}
If we use
\begin{equation} \label{e_gamma}
	\int_0^\infty dx \, x^{s-1} e^{-iB x} = B^{-s} \Gamma(s) e^{-i\pi s/2}, \quad \text{for } \text{Im}(s) > 0,
\end{equation}
We get
\begin{equation}
	P_{\text{exc}}^{\text{(der)}} = g^2\nu^2  \left| (2\omega)^{2i\nu} B^{2i\nu} \Gamma(-2i\nu) e^{-iA} e^{-\pi \nu} \right|^2.
\end{equation}
Simplifying further, and noting that the terms $(2\omega)^{2i\nu}$ and $B^{2i\nu}$ are pure phase factors with modulus equal to one, we find the exact relation
\begin{equation}
	P_{\text{exc}}^{\text{(der)}} = \tilde{g}^2 \nu^2 \cdot |\Gamma(-2i\nu)|^2 \cdot e^{-2\pi \nu},
\end{equation}
and with the use of the identity,
\begin{equation}
	|\Gamma(i x)|^2 = \frac{\pi}{x \sinh(\pi x)} \Rightarrow |\Gamma(-2i\nu)|^2 = \frac{\pi}{2\nu \sinh(2\pi \nu)},
\end{equation}
we obtain
\begin{equation}
	P_{\text{exc}}^{\text{(der)}} = \frac{\pi \tilde{g}^2 \nu e^{-2\pi \nu}}{2 \sinh(2\pi \nu)}.
\end{equation}
Or, using the identity $\sinh(2\pi \nu) = (1/2)(e^{2\pi\nu} - e^{-2\pi\nu})$, we finally obtain
\begin{equation} \label{e_Pex_der}
	P_{\text{exc}}^{\text{(der)}} = \frac{\pi\tilde{g}^2\nu}{e^{4\pi \nu} - 1},
\end{equation}
The result in Eq. \eqref{e_Pex_der} describes a thermal radiation spectrum as the denominator is characteristic of a Bose-Einstein distribution. The term in the exponential, $4\pi \nu$, corresponds to an effective temperature $T \propto 1/4\pi$, which is precisely the Hawking temperature of the EUP black hole in our scaled units. Considering the geometrized mass of the black hole, we have
\begin{equation} \label{e_Pex_der2}
    P_{\text{exc}}^{\text{(der)}} = \frac{\pi\tilde{g}^2\nu r_{\rm h}^{\rm EUP}}{(e^{4\pi \nu r_{\rm h}^{\rm EUP}} - 1)}.
\end{equation}
The result is remarkable because it shows that the detector, despite being in geodesic free-fall motion and interacting with what an observer at infinity defines as a vacuum state (the Boulware state), radiates as if it were immersed in a thermal bath at the Hawking temperature.

\begin{figure}
{\includegraphics[width=0.6\textwidth]{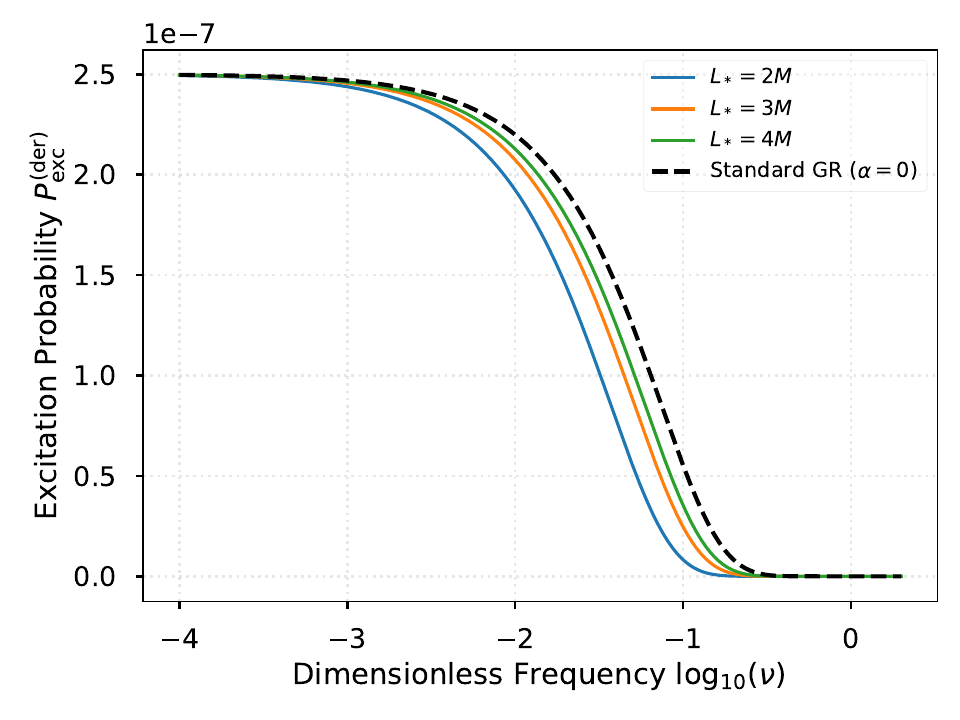}} 
    \caption{The excitation probability $P_{\text{exc}}^{\text{(der)}}$ for a derivatively coupled atom undergoing radial free-fall into a dimensionless EUP-corrected black hole ($M=1$, $\alpha=1$). The thermal radiation spectrum is plotted as a function of dimensionless frequency $\nu$ for several values of the EUP fundamental length scale, $L_*$, relative to the black hole mass. The plot shows that as $L_*$ increases, the EUP-corrected spectrum (solid lines) systematically converges to the standard GR case (dashed line), which represents the limit $\alpha = 0$. Here, $g = 0.001$.}
    \label{fig2}
\end{figure}

The results, as visualized in Figure \ref{fig2}, offer a clear and quantitative depiction of how the EUP can influence acceleration radiation. The plot confirms that the radiation spectrum remains perfectly thermal, consistent with a Bose-Einstein distribution, for all tested values of the fundamental length scale $L_*$. The primary effect of the EUP model is a modification of the black hole's effective Hawking temperature. As shown, the deviation from standard GR is most pronounced for the smallest value, $L_* = 2M$. This is the expected behavior, as the EUP correction to the black hole horizon—proportional to $(M/L_*)^2$—is largest when the new fundamental length scale is on the same order as the black hole's mass. This EUP correction leads to an enlargement of the event horizon. Since the black hole's temperature is inversely proportional to its horizon size, this enlargement results in a cooler black hole and, consequently, a less intense thermal bath for the infalling detector to interact with. This is evident in the plot, where the curve for $L_* = 2M$ shows the lowest overall probability and a thermal cutoff ("knee") at a lower frequency compared to the other cases.

Furthermore, the plot demonstrates the model's robust correspondence with GR. As the ratio $L_*/M$ increases to $3M$ and $4M$, the radiation spectrum systematically converges toward the standard GR limit (dashed line). This provides a crucial consistency check for the model and illustrates how acceleration radiation can, in principle, serve as a sensitive probe for new fundamental length scales predicted by quantum gravity theories.

Our finding for the EUP black hole is also structurally consistent with the thermal spectrum recently reported for the Schwarzschild spacetime by Das et al. \cite{Das:2025rzz}, who investigated the derivative coupling interaction in that specific geometry. The significance of our work, however, lies in the fact that this result was derived from our general expression for the excitation probability in Eq. \eqref{e_Pex}, which applies to a broad class of static, spherically symmetric spacetimes. Its successful application here to a non-trivial, quantum-corrected background confirms the robustness of the phenomenon and validates our formalism as a powerful tool for probing near-horizon physics in various theories of gravity.

Compared with the standard (non-derivative) coupling result \cite{Scully:2017utk}, our analysis reveals a significant difference in the interaction strength. In the near-horizon limit, the geometric prefactor in the probability amplitude for derivative coupling is enhanced by a factor of approximately 2 relative to the minimal coupling case. Since the probability is proportional to the amplitude squared, this leads to an enhancement of the excitation probability by a factor of four for the Schwarzschild case:
    \begin{equation} \label{e_Pex_der3}
        P_{\text{exc}}^{\text{(der)}} \approx 4 P_{\text{exc}}^{\text{(min)}}.
    \end{equation}
This enhancement is a key physical signature of the derivative coupling, indicating that a detector sensitive to field gradients interacts more strongly with the highly compressed vacuum modes near the event horizon.

It is particularly noteworthy that this thermal spectrum is recovered using the derivative coupling model. This speaks to the universality of the thermal nature of the near-horizon quantum vacuum. The thermal character is not an artifact of a specific, simplified interaction, like the minimal coupling model, but rather a fundamental feature of the spacetime geometry itself. The choice of interaction model, however, does determine the overall efficiency and characteristics of the radiation process. The appearance of $\tilde{g}$ in the numerator indicates that the probability is inversely proportional to the square of the detector's energy gap, $\omega$. This is a distinct signature of the derivative coupling, implying that for a fixed dimensionless coupling strength $g$, atoms with higher internal frequencies will experience a significantly suppressed excitation rate compared to what might be expected from minimal coupling. This dependence underscores that the detector is physically responding to field gradients, and the efficiency of this response is tied to its own internal structure.

\section{Example 2: Ricci-coupled Kalb-Ramond Bumblebee gravity with global monopole}  \label{sec5}
In our second example, we explore a black hole solution arising from a Ricci-coupled Bumblebee gravity model with a global monopole \cite{Belchior:2025xam}. Bumblebee models are effective field theories that describe spontaneous Lorentz symmetry breaking through the vacuum expectation value of a dynamic vector field. A global monopole is a spherical topological defect, characterized by a charge $\eta$, which could have formed during a symmetry-breaking phase transition in the early universe, altering the surrounding spacetime geometry by introducing a solid angle deficit. The combined effect of these phenomena modifies the standard Schwarzschild spacetime, leading to the lapse function
\begin{equation} \label{e_bumble_met}
    f(r)=\alpha-\frac{2M}{r},\quad\alpha=\frac{1-\bar{k}\eta^2}{1-\gamma},
\end{equation}
where $\gamma$ quantifies the breaking of Lorentz symmetry, $\eta$ is the global monopole charge, and $\bar{k} = 8\pi$ is the gravitational coupling constant in our chosen geometrized units. The parameter $\alpha$ thus encapsulates the deviation from the standard spacetime structure. To analyze the leading-order effects, we define a small perturbation parameter $a = \alpha - 1$. This allows us to perform a first-order Taylor expansion in $a$ to analytically solve the otherwise intractable geodesic integrals.

\begin{figure}
{\includegraphics[width=0.6\textwidth]{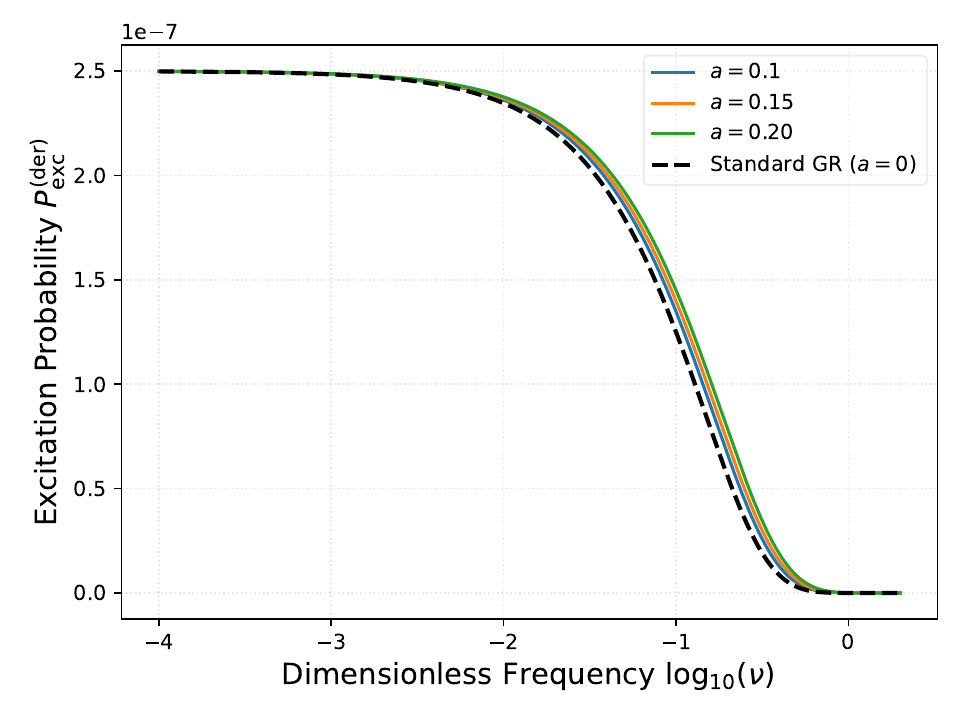}} 
    \caption{The excitation probability $P_{\text{exc}}^{\text{(der)}}$ for a derivatively coupled atom in radial free-fall, plotted as a function of dimensionless frequency $\nu$ for a black hole in a Ricci-coupled Bumblebee gravity model. The thermal spectrum is shown for several values of the Lorentz-violating parameter $a$. The results are compared to the standard GR case, which corresponds to $a = 0$. The plot demonstrates that as the parameter $a$ increases, the radiation is enhanced, and the effective temperature of the spectrum increases, shifting the thermal cutoff to higher frequencies. Here, $g = 0.001$.}
    \label{fig3}
\end{figure}

We begin with the static and spherically symmetric line element given in Eq. \eqref{e_metric}, and with Eq. \eqref{e_bumble_met}, the radius of the event horizon is found as $r_{\rm h}^{\rm BB} = 2M/\alpha$. For simplicity, we scale it to $r_{\rm h}^{\rm BB} = 1$. Furthermore, we only assume a case where $\alpha$ deviates from $1$, representing a small deviation from the EUP case. The prefactor in Eq. \eqref{e_Pex} is then
\begin{equation}
	\frac{\left(1 + \sqrt{1 - f(r)} \right)}{f(r)\sqrt{1 - f(r)}} \approx \frac{r}{\sqrt{r}-1} + \left[ \frac{r \left(r^{3/2}-3 \sqrt{r}-2\right)}{2 (r -1)} \right]a,
\end{equation}
where $a = \alpha - 1$. It would mean that if $\alpha \sim 1$, then $a \sim 0$. Next, we evaluate the equations in \eqref{e_geo} resulting to
\begin{align}
	\tau & = -\frac{2 r^{3/2}}{3}-\frac{r^{3/2} \left(3 r -5\right) a}{15} + C_1     , \nonumber \\
	t & = -\frac{2 r^{3/2}}{3} - 2\sqrt{r} + \ln{\left( \sqrt{r}+1 \right)} - \ln{\left( \sqrt{r}-1 \right)} + C_1     \nonumber \\
	&+a\left[ -\frac{r^{5/2}a}{5} +\frac{2 r^{3/2}}{3} + 2\sqrt{r} - \ln{\left( \sqrt{r}+1 \right)} + \ln{\left( \sqrt{r}-1 \right)}  \right] + C_2 , \nonumber \\
	r_* & = r + \ln\left( r - 1 \right) - a\left[ r + \ln\left( r - 1 \right) \right] + C_3   .
\end{align}
Using these, we set the excitation probability as
\begin{equation}
	P_{\text{exc}}^{\text{(der)}} = \tilde{g}^2 \nu^2 \left|\int_{1}^\infty dr \left\{\frac{r}{\sqrt{r}-1} + \left[ \frac{r \left(r^{3/2}-3 \sqrt{r}-2\right)}{2 (r -1)} \right]\right\}a e^{-i \nu \Phi(r)} e^{-i\omega \Omega(r)} \right|^2,
\end{equation}
where
\begin{align}
    \Phi(r) &= \left[ 2\sqrt{r} + r + (2/3)r^{3/2} + 2\ln\left( \sqrt{r} - 1 \right) \right]\left(1-a\right) + \frac{ar^{5/2}}{5}, \nonumber \\
    \Omega(r) &= \frac{2 r^{3/2}}{3}\left[1+\frac{ \left(3 r -5\right) a}{10}\right]
\end{align}
After implementing the following substitutions $y = r^{5/2}$, and $x = (2\omega/3)(y-1)$, we find
\begin{align}
	P_{\text{exc}}^{\text{(der)}} &= \tilde{g}^2 \nu^2 \left|\int_{0}^\infty dx\frac{(1-a)}{x}e^{-i\nu \Phi(x)e^{-i\omega\Omega(x)}}\right|^2,
\end{align}
where
\begin{align}
    \Phi(x) &= \left[\frac{11}{3} + 2\ln\left( \frac{3}{10} \right) +2\ln\left( \frac{x}{\omega}\right)\right]\left( 1-a \right) + \frac{a}{5}  + \frac{3x}{5\omega}\left( 1-\frac{a}{2} \right), \nonumber \\
    \Omega(x) &= \frac{2}{3} + \frac{3}{5}\left( \frac{x}{\omega} - \frac{2a}{3} \right).
\end{align}
After some manipulation,
\begin{equation}
    P_{\text{exc}}^{\text{(der)}} = \tilde{g}^2 \nu^2 (1-a)^2 \left|\int_{0}^\infty dx\frac{(1-a)}{x} \left( \frac{3x}{10w} \right)^{-2i\nu(1-a)} e^{-iA}e^{-ixB} \right|^2,
\end{equation}
where the phase constant $A$ and the parameter $B$ are given by
\begin{equation}
    A = \nu(1-a)\left[\frac{11}{3} + 2\ln\left(\frac{3}{10\omega}\right)\right] + \frac{a\nu}{5} + \frac{2\omega}{3}\left(1-\frac{a}{5}\right), \quad B = \frac{3}{5}\left[ 1+\frac{\nu}{\omega}\left(1-\frac{a}{2}\right) \right].
\end{equation}
The above simplifies to
\begin{equation}
    P_{\text{exc}}^{\text{(der)}} = \tilde{g}^2 \nu^2 (1-a)^2 \left|\int_{0}^\infty dx \, x^{-1-2i\nu(1-a)} \, e^{-ixB}    \right|^2.
\end{equation}
Using Eq. \eqref{e_gamma}, we obtain the expression
\begin{align}
    P_{\text{exc}}^{\text{(der)}} &= \tilde{g}^2 \nu^2 (1-a)^2 \cdot |\Gamma(-2i\nu(1-a)|^2 \cdot e^{-2\pi \nu (1-a)} \nonumber \\
    &= \frac{\pi \tilde{g}^2 \nu (1-a)e^{-2\pi \nu (1-a)}}{2 \sinh(2\pi \nu (1-a))}.
\end{align}
Finally, using the exponential identity of $\sinh(x)$,
\begin{equation} \label{e_BBPex}
    P_{\text{exc}}^{\text{(der)}} = \frac{\pi \tilde{g}^2 \nu (1-a)}{e^{4\pi \nu (1-a)}-1}.
\end{equation}
We can see that the thermal character of the acceleration radiation spectrum is preserved, while the parameters of the underlying gravity theory directly alter its properties. The term $(1-a)$ now appears in both the numerator and the exponential, signifying a direct influence of the Lorentz-violating Bumblebee field and the global monopole on the radiation process. We plot the behavior of this function as shown in Fig. \ref{fig3}. 
The influence of the Lorentz-violating parameter $a$ from the Bumblebee gravity model on the acceleration radiation spectrum is illustrated in Figure \ref{fig3}. A key result is that while the thermal character of the radiation is preserved, its properties are directly and significantly modified by the underlying gravitational theory. As the plot clearly shows, a positive, non-zero value for $a$ leads to a marked enhancement of the excitation probability. This behavior can be understood from the derived thermal spectrum, whose effective temperature scales as $T \propto 1/(1-a)$. For a small positive $a$, this results in an effective temperature that is higher than in the standard GR case (where $a=0$). This hotter thermal bath experienced by the infalling atom results in a more intense radiation flux, with a spectral peak shifted toward higher frequencies. The plot visualizes this effect, with the curve for $a = 0.3$ showing the highest intensity.

Such a result is particularly insightful when contrasted with the EUP black hole scenario discussed previously. Whereas the EUP model led to a suppression of radiation due to an effectively lower temperature, the Bumblebee model, as parameterized here, leads to its enhancement. This provides a distinct theoretical signature. It demonstrates that acceleration radiation is not only a probe for the existence of new physics beyond GR, but that its spectral characteristics—specifically, whether the effective temperature is raised or lowered—could serve as a powerful diagnostic tool to differentiate between various classes of alternative gravity theories.

To analyze the effective temperature that this spectrum implies, note that the exponential gives information about that effective temperature. That is, $T \propto 1/4\pi(1-a)$ in our scaled units. However, since $a \sim 0$, $T \propto (1+a)/4\pi$, which readily reduces to the Schwarzschild Hawking temperature if $a = 0$. Therefore, we can write Eq. \eqref{e_BBPex} as
\begin{equation}
    P_{\text{exc}}^{\text{(der)}} = \frac{\pi \tilde{g}^2 \nu r_{\rm h}^{\rm Schw} (1-a)}{\left[e^{4\pi \nu r_{\rm h}^{\rm Schw} (1-a)}-1 \right]}.
\end{equation}
The fact that the temperature derived from the detector's response perfectly matches the black hole's thermodynamic Hawking temperature to first order in $a$ is a crucial validation of the framework's self-consistency. This is a non-trivial result: the Hawking temperature is a global property of the spacetime derived from its geometric structure, while the detector's response is calculated from a local, microscopic interaction along its worldline. Their agreement confirms that acceleration radiation acts as a faithful local probe of the global thermal properties of the near-horizon quantum vacuum, even within the context of this modified gravity theory.

\section{Implications to Entropy} \label{sec6}
Suppose we write $\tilde{\nu} = \nu(1-a)$. Now, we compare Eq. \eqref{e_BBPex} with the excitation probability that used minimal coupling, 
\begin{equation}
    \frac{P_{\text{exc}}^{\text{(der)}}}{P_{\text{exc}}^{\text{(min)}}} = \frac{g^2 (1-a) (e^{4\pi \nu}-1)}{4\omega^2 (e^{4\pi \tilde{\nu}}-1)}.
\end{equation}
Such results show that the derivative coupling leads to stronger excitation at high-frequency atomic transitions, and more enhancement is achieved due to the Bumblebee parameter $\alpha$ (encapsulated in $a$). From Ref. \cite{Scully:2017utk}, the HBAR entropy flux (via the von Neumann entropy rate) is given by
\begin{equation}
    \dot{S}_p = -k_{\rm B} \, \text{Tr}(\dot{\rho} \ln \rho) \approx 4\pi k_{\rm B} r_g \sum_\nu \nu \, \dot{\bar{n}}_\nu.
\end{equation}
That is,
\begin{equation}
    \dot{S}_p \propto \sum_\nu \nu \Gamma_\nu,
\end{equation}
where $\Gamma_\nu \propto P_{\text{exc}}$ gives the rate of photon emission at frequency $\nu$. Due to $P_{\text{exc}}^{\text{(der)}}$ is enhanced compared to minimal coupling (especially for large $\omega$), the emitted flux $\dot{\bar{n}}_\nu$ increases accordingly. Since the entropy flux is linear in both $\nu$ and $\dot{\bar{n}}_\nu$, this leads to
\begin{equation}
 \dot{S}_p^{\text{(der)}} > \dot{S}_p^{\text{(min)}} \implies \frac{\dot{S}_p^{\text{(der)}}}{\dot{S}_p^{\text{(min)}}} \sim   \frac{g^2 (1-a) (e^{4\pi \nu}-1)}{4\omega^2 (e^{4\pi \tilde{\nu}}-1)},
\end{equation}
telling that atoms with high-frequency internal transitions falling into a black hole and interacting via derivative coupling emit significantly more entropy in the acceleration radiation. These results may imply that the derivative coupling picks up the gradient of the field, which becomes larger near the black hole horizon, where redshift and mode compression are so extreme. Physically, it enables probing shorter-wavelength modes, increasing the effective participation of the field vacuum fluctuations, and hence, more entropy is radiated out, which has implications for the robustness of the HBAR entropy-area correspondence, potentially providing a model-dependence correction to the entropy-area relation:
\begin{equation}
     \dot{S}_p^{\text{(der)}} = \frac{g^2 (1-a) (e^{4\pi \nu}-1)}{4\omega^2 (e^{4\pi \tilde{\nu}}-1)} \frac{\dot{A}_p}{4}.
\end{equation}
The interaction of the detector with the quantum vacuum can be understood as a form of quantum friction. As the detector accelerates relative to the field modes, it experiences a drag that excites these modes, creating thermal radiation ("heat") and increasing the entanglement between the detector and the field. The resulting growth in the von Neumann entropy of the field is a direct measure of this generated thermal disorder. Our findings indicate that this quantum friction is significantly stronger for a derivative-coupled detector. Its heightened sensitivity to field gradients allows it to interact much more efficiently with the highly blueshifted and compressed vacuum modes near the horizon. This results in a more rapid generation of thermal radiation and, consequently, a substantially larger entropy flux, indicating that the mechanism of entropy extraction from the vacuum is highly dependent on the nature of the interaction.

\section{Conclusion} \label{conc}
In this work, we have developed a general theoretical framework for investigating the acceleration radiation emitted by an Unruh-DeWitt detector with a non-minimal, derivative coupling to a massless scalar field. We considered the specific case of a detector on a purely radial, geodesic trajectory, infalling into a generic static and spherically symmetric black hole. The central achievement of our analysis is the derivation of an integral expression for the detector's excitation probability, $P_{\text{exc}}^{\text{(der)}}$. This result, obtained from first principles using time-dependent perturbation theory, explicitly incorporates the detector's four-velocity, coupling its internal state to the gradient of the quantum field as experienced along its worldline. A key feature of our final expression is its generality; the specific properties of the background spacetime are encapsulated entirely within the metric function, $f(r)$, confirming that the detector's response is dynamically influenced by the local gravitational environment.

The application of this formalism to specific spacetimes yielded significant physical insights. For the EUP black hole, we demonstrated that the detector radiates with a perfect thermal spectrum, albeit at a temperature that is suppressed by the EUP corrections. The recovery of a thermal spectrum, even with a non-minimal interaction model and in a quantum-corrected spacetime, reinforces the conclusion that the thermal nature of the near-horizon vacuum is a universal feature. More importantly, this demonstrates that the characteristics of acceleration radiation serve as a direct probe of the underlying spacetime geometry, capable of distinguishing between different theories of modified gravity. More importantly, when we extended the analysis to a black hole in a Ricci-coupled Kalb-Ramond Bumblebee gravity model, we found that the radiation spectrum was modified in direct correspondence with the theory's Lorentz-violating parameter. The effective temperature of the radiation was shown to match the modified black hole's thermodynamic Hawking temperature to first order, establishing acceleration radiation as a potential theoretical probe for deviations from GR. Furthermore, our results indicate that derivative coupling leads to an enhanced entropy flux compared to minimal coupling, suggesting a more efficient mechanism for entropy extraction from the near-horizon field gradients, akin to a stronger "quantum friction" effect.

The general framework established in this paper provides a foundation for several avenues of future inquiry. The most immediate extension is the numerical application of our expression for the excitation probability to a wider variety of black hole spacetimes, such as Reissner-Nordstr\"om or other solutions in alternative theories of gravity, to explore the distinct spectral signatures of charge or other physical parameters. Further theoretical development could proceed by relaxing the simplifying assumptions of the present model. For instance, the analysis could be extended to detectors on more complex, non-geodesic trajectories or those possessing orbital angular momentum, which is particularly relevant for astrophysical scenarios. Finally, investigating the model's dependence on the type of quantum field (e.g., vector or spinor fields) or moving beyond the first-order perturbative approximation would provide a more complete picture of the quantum processes at play near a black hole event horizon.

\acknowledgments
R. P. and A. \"O. would like to acknowledge networking support of the COST Action CA21106 - COSMIC WISPers in the Dark Universe: Theory, astrophysics and experiments (CosmicWISPers), the COST Action CA22113 - Fundamental challenges in theoretical physics (THEORY-CHALLENGES), the COST Action CA21136 - Addressing observational tensions in cosmology with systematics and fundamental physics (CosmoVerse), the COST Action CA23130 - Bridging high and low energies in search of quantum gravity (BridgeQG), and the COST Action CA23115 - Relativistic Quantum Information (RQI) funded by COST (European Cooperation in Science and Technology). A. \"O. also thanks to EMU, TUBITAK, ULAKBIM (Turkiye) and SCOAP3 (Switzerland) for their support.

\bibliography{ref}

\end{document}